# Accurate prediction of the properties of materials using the CAM-B3LYP Density Functional

*Musen Li,[1] Jeffrey R. Reimers,[1,2]\* Michael J. Ford,[2] Rika Kobayashi,[3] and Roger D. Amos[3]*

1 International Centre for Quantum and Molecular Structures and Department of Physics, Shanghai University, Shanghai 200444, China.

2 University of Technology Sydney, School of Mathematical and Physical Sciences, Ultimo, New South Wales 2007, Australia.

3 ANU Supercomputer Facility, Leonard Huxley Bldg. 56, Mills Rd, Canberra, ACT, 2601, Australia.

Email: jeffrey.reimers@uts.edu.au, Mike.Ford@uts.edu.au, Rika.Kobayashi@anu.edu.au, Roger.Amos@anu.edu.au

**Abstract**

Density functionals with asymptotic corrections to the long-range potential provide entry-level methods for calculations on molecules that can sustain charge transfer, but similar applications in Materials Science are rare. We describe an implementation of the CAM-B3LYP range-separated functional within the Vienna Ab-initio Simulation Package (VASP) framework, together with its analytical functional derivatives. Results obtained for eight representative materials: aluminum, diamond, graphene, silicon, NaCl, MgO, 2D h-BN and 3D h-BN, indicate that CAM-B3LYP predictions embody mean-absolute deviations (MAD) compared to HSE06 that are reduced by a factor of 6 for lattice parameters, 4 for quasiparticle band gaps, 3 for the lowest optical excitation energies, and 6 for exciton binding energies. Further, CAM-B3LYP appears competitive compared to ab initio $G_0W_0$ and Bethe-Salpeter equation (BSE) approaches. The CAM-B3LYP implementation in VASP was verified by comparison of optimized geometries and reaction energies for isolated molecules taken from the ACCDB database, evaluated in large periodic unit cells, to analogous results obtained using Gaussian basis sets. Using standard GW pseudopotentials and energy cutoffs for the plane-wave calculations and the aug-cc-pV5Z basis set for the atomic-basis ones, the MAD in energy for 1738 chemical reactions was 0.34 kcal mol$^{-1}$, whilst for 480 unique bond lengths this was 0.0036 Å; these values reduced to 0.28 kcal mol$^{-1}$ (largest error 0.94 kcal mol$^{-1}$) and 0.0009 Å by increasing the plane-wave cuttoff energy to 850 eV.

**Keywords**

Density-functional theory, Materials science, Plane-wave basis set, Long-range asymptotic potential error, Quasiparticle bandgap, Optical transition energies in materials, Exciton binding energy

# 1. Introduction

There are two significant shortcomings in density functional theory (DFT)[1,2] approaches that attempt to treat all molecules and materials starting from the description of the homogeneous electron gas, including both the local density approximation[3] (LDA) functional, all general gradient approximation (GGA)[4] functionals, and also hybrid approaches.[5] Of these, one is associated with missing non-dynamic electron correlation,[6] which can manifest severely in that many possible electronic states do not satisfy either the Kohn-Sham[1] or Gunnarsson-Lundqvist[7] theorems and hence cannot be described using DFT. The other is the self-interaction error (SIE)[8] that arises as the inexact density functionals in common use fail to properly balance electron correlation and exchange effects, incorrectly allowing each electron to interact with itself. In this work, we implement, test, and apply a method designed to mitigate the effects of SIE into our personal copy of the Vienna Ab-initio Simulation Package (VASP)[9,10] version 5.4.4. This package is one that is widely applied for electronic-structure modelling in solid-state physics.

A critical aspect of the SIE is that the potential energy of any electron, when removed to infinity, is incorrect. Savin and co-workers[11,12] noted that *ab initio* electronic structure approached based on Hartree-Fock[13] theory did not suffer from this problem as, at all levels of the theory, balance between exchange and correlation is guaranteed. As such approaches, when implemented at computationally expedient low-levels, do not describe critical short-range effects as well as do density functional approaches, they suggested use of range separated DFT (rs-DFT) in which hybrid functionals that include aspects of both LDA and HF exchange are modified so that only the HF component is applied asymptotically. This led to the development of long-range-corrections to DFT (LC-DFT) functionals by Hirao and coworkers[14] that were designed to mitigate the SIE. Soon thereafter, LC-DFT was shown to deliver good performance for molecule polarizabilities, charge-transfer excitations, nonlinear optical properties, magnetic properties, and bond dissociation energies.[15-17] In the initial treatments, the contribution of HF exchange was dampened to zero at short range, a feature that counteracts other advantages of hybrid functionals, leading Yanai and co-workers[18] to suggest a variation, the Coulomb-attenuated B3LYP (CAM-B3LYP) method in which the damping function was empirically optimized to deliver robust performance for a wider range of molecular properties.

The CAM-B3LYP density functional was quickly introduced[19] into the Gaussian-03 package (and its descendants) that is widely used for modelling molecular properties, showing excellent results. Then it was demonstrated to be able to describe accurately charge-transfer transitions in large aromatic molecules[20] for which conventional LDA, GGA, and hybrid functionals perform very poorly.[21] This opened up the fields of natural and artificial photosynthesis, including exciton transport and splitting, to first-principles calculations.[22-24] It is also known as the only method established to usefully predict high-resolution Huang-Rhys factors, and hence electronic spectra, in such systems.[25-27]

The effect has been broadly recognized as applying to all applications of DFT, including both molecular and periodic systems.[28] This includes the nanophotonics research field[29-33] in which optically active components are imbedded as defects into solid-state environments, with applications ranging from from quantum networks and quantum information to spin-photon interfaces.[33-36] Whereas research on defects in diamond and silicon carbide has historically been prominent,[35,37] of immediate concern are defects in hexagonal boron nitride (h-BN), which can display single-photon emission[38-41] (SPE) and optically detected magnetic resonance (ODMR).[42,43] Calculations have been essential in the determination of the chemical nature of defects and their intricate magnetic, photophysical and photochemical properties,[44-48] with ones successful in spectroscopic modelling using advanced methods such as CAM-B3LYP and or *ab initio* coupled-cluster approaches[45,46,48] or else mixed DFT and *ab initio* methods.[47] Even though calculations on localized-defect transitions using molecular models converge very quickly with increasing sample size[49] to the (more slowly converging) results from 2D simulations, charge-transfer properties can be of interest that can only be modelled using periodic models. Such problems can only be investigated by DFT using asymptotically corrected density functionals implemented into periodic codes.

Traditionally, however, the most commonly applied DFT approaches are either GGA functionals such as PBE[4] or, when needed, the screened hybrid functional HSE06.[50,51] Neither of these methods embody asymptotic correction of the SIE. These and similar functionals give poor performance for defects in general,[46,52] and can fail dramatically if charge transfer is involved.[53] Note that, whilst HSE06 is a range-separated hybrid, the range-separation parameters are chosen to maximize computational expedience and optimize metallic properties rather than to correct the SIE, making it more like a GGA functional when it comes to understanding transitions involving localized defect orbitals and conduction/valence band orbitals.

The first relevant implementation of CAM-B3LYP into a code for materials-science modelling has just been reported,[54] an implementation into the Car-Parrinello Molecule Dynamic (CPMD) package for gamma-point-only calculations. The results obtained demonstrated enhanced optical spectra prediction, as expected.

In VASP, as in many other programmes that implement plan-wave basis sets, the HF exact-exchange interactions are calculated in reciprocal-space.[55,56] The rs-DFT functional HSE06 is already implemented in VASP and many other codes, and, implementation of CAM-B3LYP would seem to require a straightforward modification of the existing code. Unfortunately, there are subtle differences between HSE06 and CAM-B3LYP that need to be taken into account. For HSE06, an approximate treatment is defined in the functional[57] concerning how the range separation influences functional evaluation. This allows simple code to be written for HSE06, but no such approximation is included in the definition of CAM-B3LYP.[18] We implement the full CAM-B3LYP functional, as well as its approximation utilizing an expedient method implementation based upon the existing implementation of HSE06. Variations[58,59] of the basic functional may be selected using input parameters.

The developed code is tested by optimizing geometries and calculating reaction energies for reference molecules and comparing the results to those analogously obtained using Gaussian-16.[60] Various atomic basis sets were used in the Gaussian-16 calculations, seeking convergence towards the complete basis set limit. Similarly, large plane-wave basis sets were used in the VASP calculations, as well as large periodic boxes to minimize intermolecular interactions. Various available pseudopotentials were used to describe the core electrons in the VASP calculations.

The molecules and reactions were taken from A Collection of Chemistry Databases (ACCDB).[61] This collect together six popular benchmark databases: MGCDB84,[62] GMTKN55,[63] Minnesota Database 2015B,[64] DP284,[65] Metals & EE,[66] and W4-17.[67] This collection includes a wider variety of applications and reflects benchmark results more reliable than its alternatives, including, e.g., Pople's G2-1 dataset.[68] In total, 203 molecular structures were optimized and 1738 chemical reactions considered that involved H, Be, B, C, N, O, F, Al, Si, P, S, and Cl atoms.

Then some basic applications to materials science are developed. We consider calculated lattice vectors, quasiparticle band gaps, lowest optical excitation energies, and exciton binding energies for the representative materials: aluminum, graphene, diamond, silicon, NaCl, MgO, and h-BN (both 2D and 3D), comparing the results to experiment, $G_0W_0$, this used with also the Bethe-Salpeter equation (BSE) to calculate optical transition energies, and with results from HSE06 calculations.

## 2. Methods

*2.1 Integrals required for evaluation.*

In VASP, DFT schemes are most usually implemented in association with the projector augmented-wave (PAW) method[69] that facilitates the implicit treatment of core electrons. In the absence of an external field, the total electronic energy $E[\rho]$ as a functional of the electron density $\rho$ is written as

$$E[\rho] = T_s[\rho] + E_{xc}[\rho] + U_{Hartree}[\rho], \qquad (1)$$

where the kinetic energy is $T_s[\rho]$, the Hartree energy is $U_{Hartree}[\rho]$, and exchange-correlation energy is $E_{xc}[\rho]$. A most common practice, taken throughout this work, is to simplify the exchange-correlation functional by separating it into two additive contributions,

$$E_{xc}(\rho) = E_x(\rho) + E_c(\rho) \qquad (2)$$

depicting exchange, $E_x(\rho)$, and correlation $E_c(\rho)$. To evaluate Eqn. (1), the electron density is manifested internally in equivalent forms as a function of either position-space Cartesian coordinates **r**, $\rho(\mathbf{r})$, or else in the associated momentum-space representation $\rho(\mathbf{p})$, evaluated by summing over all $N$ occupied 1-particle orbital densities as

$$\rho(\mathbf{r}) = \sum_i^N |\psi_i(\mathbf{r})|^2, \quad \rho(\mathbf{p}) = \sum_i^N |\psi_i(\mathbf{p})|^2, \qquad (3)$$

where the orbital wavefunctions $\psi_i(\mathbf{r})$ and $\psi_i(\mathbf{p})$ are related by Fourier transformation and obtained by solving the Kohn-Sham equation [1], which, e.g., in position space, is

$$\left(-\frac{1}{2}\Delta + V_H[\rho(\mathbf{r})] + V_{xc}[\rho(\mathbf{r})]\right)\psi_i(\mathbf{r}) = \epsilon_i \psi_i(\mathbf{r}) . \qquad (4)$$

Note that Eqn. (3) involves a sum over all electrons. In spin-polarized calculations, this can be conveniently divided into sums over electrons of each spin type $\sigma$; we show this separation in the subsequent equations only when necessary.

For the B3LYP exchange-correlation hybrid density functional,[5] the exchange functional $E_{xB3LYP}$ is expressed as a linear combination of the Hartree-Fock exchange operator[13] $E_{xHF}$, the Becke88 GGA exchange functional[70] $E_{xB88}$, and the LDA exchange functional[3] $E_{xLDA}$, as

$$E_{xB3LYP} = 0.2 E_{xHF} + 0.8 E_{xLDA} + 0.72 \Delta E_{xB88} \qquad (5)$$

where $\Delta E_{xB88} = E_{xB88} - E_{xLDA}$ is the gradient correction introduced to the exchange as a result of the GGA expansion of the LDA. Its analogous correlation functional cannot be so separated, and hence is expressed as a sum of the LYP GGA-correlation functional[71] $E_{cLYP}$ and the LDA correlation functional[3] $E_{cLDA}$, via

$$E_{cB3LYP} = 0.19 E_{cLDA} + 0.81 E_{cLYP}. \qquad (6)$$

Yanai et al.[18] modified B3LYP to create CAM-B3LYP by modifying the exchange interaction to mix its constituents parts according to the distance $r_{12} = |\mathbf{r_2} - \mathbf{r_1}|$ between two points in space, identifying a short-range (SR) part and a long-range (LR) part, rewriting

$$\frac{1}{r_{12}} = \frac{[\alpha + \beta \text{erf}(\mu r_{12})]}{r_{12}} + \frac{1 - [\alpha + \beta \text{erf}(\mu r_{12})]}{r_{12}}, \quad (7)$$

where $\text{erf}(\mu r_{12})$ is the error function

$$\text{erf}(\mu r_{12}) = 1 - \frac{2}{\sqrt{\pi}} \int_{\mu r_{12}}^{\infty} e^{-t^2} dt . \quad (8)$$

and $\mu$ defines the region of range separation, being inversely proportional to the short-range extent. The first-term in Eqn. (7) is used to define LR contributions $E_x^{LR}$ to the total exchange $E_x$, whereas the second term is used to define SR contributions $E_x^{SR}$. The operators whose expectation values lead to the electronic energy then, for CAM-B3LYP, become

$$\hat{E}_x = \hat{E}_x^{LR} + \hat{E}_x^{SR} = \frac{[\alpha + \beta \text{erf}(\mu r_{12})]}{r_{12}} \hat{E}_{xHF} + \frac{1 - [\alpha + \beta \text{erf}(\mu r_{12})]}{r_{12}} \hat{E}_{xB88} \quad (9)$$

A subtle feature of rc-DFT is that the exchange energy can no longer be separated into individual terms corresponding to LDA and post-LDA contributions, like is done in Eqn. (5) for B3LYP. For CAM-B3LYP, this means that $E_{xB88}$ can no longer be expressed as sums of terms $E_{xLDA} + \Delta E_{xB88}$. From Eqn. (9), the CAM-B3LYP exchange energy can be expressed in terms of the original energies and either the SR or LR energies as

$$E_{xCAM-B3LYP} = \alpha E_{xHF} + \beta E_{xHF}^{LR} + (1 - \alpha) E_{xB88} - \beta E_{xB88}^{LR} \quad (10)$$

$$= (\alpha + \beta) E_{xHF} - \beta E_{xHF}^{SR} + (1 - \alpha - \beta) E_{xB88} + \beta E_{xB88}^{SR}. \quad (11)$$

The standard ($E_{xHF}$) and long range Hartree-Fock exchange integrals [56] ($E_{xHF}^{LR}$) are:

$$E_{xHF} = -\frac{e^2}{2} \sum_{nk,mq} f_{nk} f_{mq} \iint d^3\mathbf{r} d^3\mathbf{r'} \frac{1}{\Delta r} \psi_{nk}^*(\mathbf{r}) \psi_{mq}^*(\mathbf{r'}) \psi_{nk}(\mathbf{r}) \psi_{mq}(\mathbf{r'}) \quad (12)$$

and

$$E_{xHF}^{LR} = -\frac{e^2}{2} \sum_{nk,mq} f_{nk} f_{mq} \iint d^3\mathbf{r} d^3\mathbf{r'} \frac{\text{erf}(\mu \Delta r)}{\Delta r} \psi_{nk}^*(\mathbf{r}) \psi_{mq}^*(\mathbf{r'}) \psi_{nk}(\mathbf{r}) \psi_{mq}(\mathbf{r'}), \quad (13)$$

where $\mathbf{r}$ and $\mathbf{r'}$ are points in coordinate space separated by a distance $\Delta r$, $\psi$ are the occupied molecular or crystal orbitals indexed by the band numbers $n$ and $m$ as well as the $k$-points indices $k$ and $q$, and $f$ are sets of occupational numbers. If $\rho_\sigma(\mathbf{r})$ and $\Delta \rho_\sigma(\mathbf{r})$ are the electron densities and density gradients for each spin $\sigma$ at all points $\mathbf{r}$ in space, then the short-range total exchange contribution, including explicit summation over spin densities, is[14]

$$E_{xB88}^{SR} = \sum_\sigma E_{xB88\sigma}^{SR} = -\frac{1}{2} \sum_\sigma \int_{-\infty}^{\infty} d^3\mathbf{r}\, \rho_\sigma^{4/3}(\mathbf{r})\, K_{B88\sigma}(\mathbf{r})\, H_{B88\sigma}(\mathbf{r}) , \quad (14)$$

where

$$H_{B88\sigma}(\mathbf{r}) = 1 - \frac{8}{3} A_\sigma(\mathbf{r}) \left[ \pi^{1/2} \text{erf} \frac{1}{2A_\sigma(\mathbf{r})} + 2A_\sigma(\mathbf{r})(1 - A_\sigma^2(\mathbf{r})) e^{-A_\sigma^{-2}(\mathbf{r})/4} - 3A_\sigma(\mathbf{r}) + 4A_\sigma^3(\mathbf{r}) \right] ,$$

(15)

$$A_\sigma(\mathbf{r}) = \frac{\mu \sqrt{K_{B88\sigma}(\mathbf{r})}}{6\sqrt{\pi} \sqrt[3]{\rho_\sigma(\mathbf{r})}}, \tag{16}$$

$$K_{B88\sigma}(\mathbf{r}) = K_{LDA} + \Delta K_{B88\sigma}(\mathbf{r}), \tag{17}$$

$$\Delta K_{B88\sigma}(\mathbf{r}) = \frac{2\beta_{B88} X(\mathbf{r})^2}{1 + 6\beta_{B88} X(\mathbf{r}) \operatorname{arcsinh} X(\mathbf{r})}, \tag{18}$$

$$K_{LDA} = \sqrt[3]{\frac{81}{4\pi}}, \quad X(\mathbf{r}) = \frac{|\Delta \rho(\mathbf{r})|}{\rho(\mathbf{r})^{4/3}}, \quad \text{and} \quad \beta_{B88} = 0.0042. \tag{19}$$

An approximation to these equations is possible that expedites code generation. This approximation is explicitly specified in the definition of the HSE06 density functional but not for CAM-B3LYP. Hence a significant difference arises pertaining to the coding of CAM-B3LYP compared to HSE06. In the approximation, Eqn. (17) is split into two parts and the cross-terms between them that appear in Eqns. (14) – (16) ignored. This leads to

$$E_{xB88}^{SR\prime} = E_{xLDA}^{SR} + \Delta E_{xB88}^{SR}, \tag{20}$$

as an approximation to $E_{xB88}^{SR}$, where

$$E_{xLDA}^{SR} = \sum_\sigma E_{xLDA\sigma}^{SR} = -\frac{1}{2}\sum_\sigma \int_{-\infty}^{\infty} d^3\mathbf{r}\, \rho_\sigma^{4/3}(\mathbf{r})\, K_{LDA}\, H_{LDA\sigma}(\mathbf{r}), \tag{21}$$

is evaluated during the processing of LDA-only quantities, and

$$\Delta E_{xB88}^{SR} = \sum_\sigma \Delta E_{xB88\sigma}^{SR} = -\frac{1}{2}\sum_\sigma \int_{-\infty}^{\infty} d^3\mathbf{r}\, \rho_\sigma^{4/3}(\mathbf{r})\, (K_{B88\sigma}(\mathbf{r}) - K_{LDA})\, H_{B88\sigma}(\mathbf{r}), \tag{22}$$

is evaluated during the processing of GGA-only quantities. We proceed by evaluating both the exact (Eqn. (14)) and approximate (Eqn. (20)) schemes so as to ascertain the significance of their differences.

*2.2 VASP implementations of the exact and approximate exchange functionals.*

Implementation of CAM-B3LYP into VASP requires implementation of the exchange and correlation functionals. This is straightforward for the correlation functional as it is the same as that for B3LYP (Eqn. (6)) and only requires the appropriate weighting of already available $E_{cLDA}$ and $E_{cLYP}$ terms. We hence focus only on the exchange contribution.

As part of the PAW procedure, exchange integrals need to be performed over atomic centres that are expanded in terms of atomic functions with increasing angular momentum, truncating the expansion at angular momentum $L$. This value needs to be larger when long-range Hartree-Fock exchange is included than it could be otherwise, with a value of at least $L = 4$ is found to be needed for atoms with $d$ electrons. For CAM-B3LYP, we truncate the integrals at this level.

In VASP, various choices are available for the specification of the PAW pseudopotentials. Two generally available choices are the use of pseudopotentials designed to optimize DFT calculations using PBE, and those optimized to perform GW calculations. Hybrid density functionals that contain some component of $E_{xHF}$ are believed to be better treated by the GW pseudopotentials as they provide for enhanced descriptions of the role of unoccupied orbitals.[72] To test this hypothesis, we compare results obtained using both sets of pseudopotentials.

*2.3 Analytical functional for the CAM-B3LYP exchange and correlation potentials.*

Currently, VASP uses numerical derivatives to evaluate the exchange and correlation potentials for the B3LYP functional. We implement analytical equations, see Supplementary Material (SM) to replace those numerical derivatives, specifying:

$$H'_{B88\rho_\sigma} = \frac{\partial H_{B88\sigma}}{\partial A_{\rho_\sigma}} \frac{\partial A_{\rho_\sigma}}{\partial \rho_\sigma}, \tag{20}$$

$$H'_{B88\Delta\rho_\sigma} = \frac{\partial H_{B88\sigma}}{\partial A_{\rho_\sigma}} \frac{\partial A_{\rho_\sigma}}{\partial \Delta\rho_\sigma}, \tag{21}$$

$$V_{xB88\rho_\sigma} = \frac{\partial E_{xB88\sigma}}{\partial \rho_\sigma}, \tag{22}$$

$$V_{xB88\Delta\rho_\sigma} = \frac{\partial E_{xB88\sigma}}{\partial \Delta\rho_\sigma}, \tag{23}$$

$$V^{SR}_{xB88\rho_\sigma} = \frac{\partial E^{SR}_{xB88\sigma}}{\partial \rho_\sigma} = H_{B88\sigma} V_{xB88\rho_\sigma} + H'_{B88\rho_\sigma} E_{xB88\sigma}, \text{ and} \tag{24}$$

$$V^{SR}_{xB88\Delta\rho_\sigma} = \frac{\partial E^{SR}_{xB88\sigma}}{\partial \Delta\rho_\sigma} = H_{B88\sigma} V_{xB88\Delta\rho_\sigma} + H'_{B88\Delta\rho_\sigma} E_{xB88\sigma}. \tag{25}$$

*2.4 Computational Details.*

For the VASP calculations on isolated molecules, we used a 20 × 20 × 20 Å³ orthogonal unit cell at the Γ-point of the Brillion zone. Except where noted, all VASP calculations used the default plane-wave energy cutoff "ENCUT" as well as "PREC=HIGH" and either "GW" or "PBE" PAW pseudopotentials. The energy convergence criterion was set to at least $10^{-6}$ eV and geometry optimizations were terminated when the gradient fell below $10^{-5}$ eV/Å. All Gasussian-16 calculations, including geometry optimizations and energy calculations, applied CAM-B3LYP using either the 6-311+G*,[73] aug-cc-pVDZ,[74] aug-cc-pVTZ,[74] aug-cc-pVQZ,[74] and aug-cc-pV5Z[74] basis sets. Optimized Cartesian coordinates and energies for all molecules, obtained using aug-cc-pV5Z or the GW pseudopotentials, are also provided in SM.

Orbital bandgaps for materials were evaluated by expanding the *k*-point mesh systematically, seeking grid-points that maximized the highest-occupied molecular orbital (HOMO) energy and minimized the lowest-unoccupied molecular-orbital (LUMO) energy. This was found to occur when using grids of 8 × 8 × 8 for diamond, NaCl, and MgO, 12 × 12 × 12 for silicon, 18 × 18 × 1 for 2D h-BN, and 9 × 9 × 3 for 3D h-BN; see SM for full information. The lowest-excitation energies were calculated by transferring an electron from the HOMO to the LUMO, changing the net band occupancies for transitions involving indirect band gaps. The state so-produced is not an eigenfunction of electronic spin, being considerable as an equal mixture of the lowest singlet and triplet excited states. As all relevant singlet-triplet splittings are expected to be very small, this quantity should provide an accurate prediction of the lowest singlet vertical excitation energy. Note that, when using VASP in this fashion, the lowest excitation energy is obtained by scaling the resulting energy difference by the total number of $k$-points used in the calculation. As this could be a large number, extremely tight energy convergence is therefore required in the VASP calculations. Results obtained in this way were found to be numerically equivalent to those obtained from much more expensive calculations in which a large supercell is used at its Γ-point (i.e., using a $n \times n \times n$ supercell instead of $n \times n \times n$ $k$-points), see SM. Coordinates, $k$-points, basis set information, and calculated CAM-B3LYP ground-state energies are listed for all materials in SM.

# 3. Results

## 3.1 Comparison of range-corrected functionals.

The parameters $\alpha, \beta$, and µ from Eqn. (7) optimized for CAM-B3LYP[18] are listed in Table 1, along with those for other popular range-corrected density functionals including HSE06 and LC-PBE. To highlight the effects of range separation, we plot the proportion of the exchange attributable to the Hartree-Fock contribution, $E_{xHF}/E_x$, against the separation distance in Fig. 1 for these and for the conventional hybrid functional B3LYP. Conventional hybrids have a fixed ratio of $E_{xHF}$ to $E_x$ that range separation allows to vary. For LC-PBE, the relative contribution of $E_{xHF}$ varies from nothing at short range to everything at long range, eliminating all errors in the long-range potential arising from the SIE.[14,17] On the other hand, for HSE06 this contribution varies in the opposite way: the $E_{xHF}$ part is damped to zero at long-range, which make it more like its constituent GGA functional at long-range, hence maintaining its large asymptotic potential error. The complete damping out of $E_{xHF}$ at short range in LC-PBE is believed to lead to poor predictions for other properties, however.[18] Hence retention of some fraction of $E_{xHF}$ contributions at short range, combined with optimization of the long-range component, was designed for CAM-B3LYP in order to create a robust functional.[18]

**Table 1.** Values of the range-separation parameters used in various rc-DFT functionals.

|  | $\alpha$ | $\alpha + \beta$ | $\mu$ (Å$^{-1}$) |
|---|---|---|---|
| CAM-B3LYP | 0.19 | 0.65 | 0.63 |
| HSE06 specification | 0.25 | 0 | 0.200436 |
| HSE06 in G16 | 0.25 | 0 | 0.20787 |
| HSE06 in VASP | 0.25 | 0 | 0.2 |
| LC-PBE | 0 | 1 | 0.25 |

In terms of HSE06 implementational differences, Table 1 notes that the value used for $\mu$ by Gaussian-16 is 0.11 a.u.$^{-1}$, slightly different from that originally specified,[50] 0.15/2$^{1/2}$ ~ 0.106066 a.u.$^{-1}$ ~ 0.200436 Å$^{-1}$). Also in VASP, the parameter $\mu$ is also slightly different again, 0.2 Å$^{-1}$. Note that both VASP and Gaussian-16 implement the approximate formula, Eqn. (20), for the short-range exchange, according to the specifications of HSE06.[50,51]

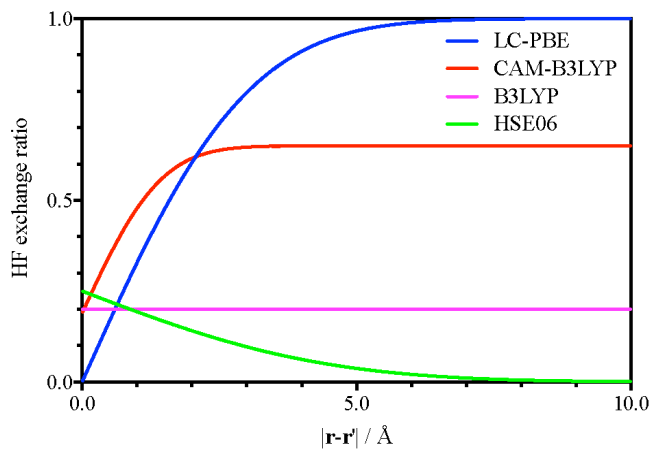

**Fig. 1.** The proportion of Hartree-Fock exchange, $E_{xHF}/E_x$, found in common range-separated density functionals as a function of the electron-electron distance, see Table 1 and Eqn. (7).

*3.2 Convergence and comparison of plane-wave versus atomic basis set calculations.*

For the 480 unique bond lengths and 1738 reaction energies from the ACCDB database, Table 2 lists the mean absolute deviation (MAD), mean relative error (MRE), and the maximum (worst-case) errors found when comparing results from VASP using plane-wave basis sets and Gaussian-16 using atomic basis sets. Five atomic basis sets are considered, as well as GW and PBE pseudopotentials for the plane-wave calculations. Detailed results for the aug-cc-pV5Z and GW approaches are provided in SM.

**Table 2.** Comparison of 480 unique bond lengths and 1738 reaction energies (for details, see SM).

| plane wave | atomic basis set | Bond lengths (Å) | | | Reaction energies (kcal mol$^{-1}$)[a] | | |
|---|---|---|---|---|---|---|---|
| PAW | Basis | MAD | MRE | Max | MAD | MRE | Max. |
| PBE | 6-311+G* | 0.0045 | -0.0035 | 0.0122 | 0.68 | 0.04 | 1.45 |
|  | aug-cc-pVDZ | 0.0092 | -0.0060 | 0.0171 | 1.13 | -0.46 | 2.90 |
|  | aug-cc-pVTZ | 0.0037 | 0.0018 | 0.0096 | 0.52 | 0.02 | 1.35 |
|  | aug-cc-pVQZ | 0.0042 | 0.0028 | 0.0104 | 0.53 | -0.05 | 2.67 |
|  | aug-cc-pV5Z | 0.0048 | 0.0032 | 0.0110 | 0.59 | -0.08 | 3.01 |
| GW | 6-311+G* | 0.0047 | -0.0037 | 0.0103 | 0.76 | 0.02 | 2.79 |
|  | aug-cc-pVDZ | 0.0097 | -0.0063 | 0.0137 | 1.11 | -0.48 | 2.98 |
|  | aug-cc-pVTZ | 0.0012 | 0.0008 | 0.0089 | 0.27 | 0.00 | 0.86 |
|  | aug-cc-pVQZ | 0.0028 | 0.0025 | 0.0092 | 0.31 | -0.07 | 1.55 |
|  | aug-cc-pV5Z | 0.0036 | 0.0027 | 0.0101 | 0.34 | -0.09 | 1.88 |
| GW 850 eV | aug-cc-pV5Z | 0.0009 | -0.0006 | 0.0061 | 0.28 | 0.02 | 0.94 |

a: 1 eV = 23.06 kcal mol$^{-1}$; 2 kcal mol$^{-1}$ is regarded is useful "chemical accuracy".

Using atomic basis sets, calculated bond lengths appear to converge quickly (Table 2), with differences to the GW results changing by only 0.0008 Å (MAD), 0.0002 Å (MRE), and 0.0009 Å (maximum), in expanding from the aug-cc-pVQZ basis set to aug-cc-pV5Z. Similarly, calculated energies change by only 0.03 kcal mol$^{-1}$ (MAD), 0.01 kcal mol$^{-1}$ (MRE), and 0.33 kcal mol$^{-1}$ (maximum).

The differences between results using the GW and PBE pseudopotentials (Table 2) are mostly small, with those obtained using GW being always the closest to the best results obtained using atomic basis sets. That the GW results are better is expected based on its improved treatment of unoccupied orbitals, but the additional computational expense may not be justified in many practical applications. The GW bond lengths differ from those obtained using aug-cc-pV5Z by only 0.0036 Å (MAD), 0.0027 Å (MRE), and 0.0101 Å (maximum); the corresponding energies differ by 0.34 kcal mol$^{-1}$ (MAD), −0.09 kcal mol$^{-1}$ (MRE), and 1.88 kcal mol$^{-1}$ (maximum).

Considering the worst-case results presented in Table 2, we note that, for the calculated bonds lengths, these all occur for the molecule SiF$_4$. Indeed, many of the poorest results occur for molecules involving fluorine. This is an extremely electron-dense atom and therefore presents the greatest challenges to accurate computation using plane-wave techniques. In Table 3, we consider the worst-case errors as a function of improvements in the plane-wave basis set associated with increasing the maximum-plane-wave-energy cut-off ENCUT. Increasing ENCUT from 550 eV to 850 eV reduces the worst-case errors in bond lengths by a factor of three. Results using GW pseudopotentials and ENCUT= 850 eV are compared to aug-cc-pV5Z ones in Table 2 for all molecules considered. The worst-case difference in both bond length and energy halve to 0.0061 Å and 0.94 kcal mol$^{-1}$, respectively. Hence the results from the plane-wave and atomic-basis-set calculations appear to be converging to high accuracy.

**Table 3.** Improvement of the MAD for the worst-case molecule pertaining to calculated bond lengths, SiF$_4$, as a function of ENCUT.

| ENCUT (eV) | MAD (Å) |
|---|---|
| 850 | 0.0033 |
| 650 | 0.0086 |
| 550 | 0.0089 |

*3.3 Cost saving in a materials application owing to the use of analytical derivatives.*

The computational scheme using analytical derivatives is expected to be more efficient, of interest to code development is by how much, given the importance of other steps in the algorithm in determining overall performance. Table 4 reports the relative computational cost for two test systems: a charged defect in 2D hexagonal boron nitride (h-BN) with one nitrogen atom missing ($V_N^{-1}$) containing 380 electrons, evaluated at the Γ-point, and a 4-atom translationally periodic silicon-crystal unit cell containing 32 electrons that was evaluated using 64 *k*-points. The ratio of the time required to complete a single cycle of the electronic self-consistent-field (SCF) procedure in each case is found to be near 0.74, indicating that the analytical method is usefully faster. In addition, Table 6 also reports the ratio of the number of cycles needed for SCF convergence, which is near 0.78, making for a total time ratio of near 0.57. The analytical derivatives are therefore of significant value.

**Table 4.** Ratio of computer resources needed using analytical derivatives compared to numerical derivatives.

| System | Time per SCF cycle | Number of SCF cycles required | Required cpu time |
|---|---|---|---|
| 380 electrons, Γ-point | 0.73 | 0.78 | 0.58 |
| 32 electrons, 64 *k*-points | 0.75 | 0.77 | 0.56 |

*3.4 Exact versus approximate short-range exchange contribution.*

In Fig. 2(a), the error introduced into the short-range exchange contribution by the replacement of Eqn. (14) with Eqn. (20), $\Delta E = |E_{xB88}^{SR\prime} - E_{xB88}^{SR}|$, is shown as a function of the electron density $\rho$ and its gradient $\Delta\rho$. The error vanishes as the density or the density gradient becomes small, as well as when the density becomes large. Hence the error is largest at some density that depends on the density gradient. This is highlighted in Fig. 2(b) where the error is shown at $\Delta\rho = 0.1$ Å$^{-4}$ as a function of the Wigner-Seitz radius $r_s = \left(\frac{3\pi}{4\rho}\right)^{1/3}$, presenting a maximum at $r_s = 2.2$ Å. In typical DFT calculations, such densities take on significant roles and hence the approximation leads to a noteworthy error, overestimating the exchange energy. This could influence singlet-triplet gaps, ferromagnetism versus antiferromagnetism, etc.

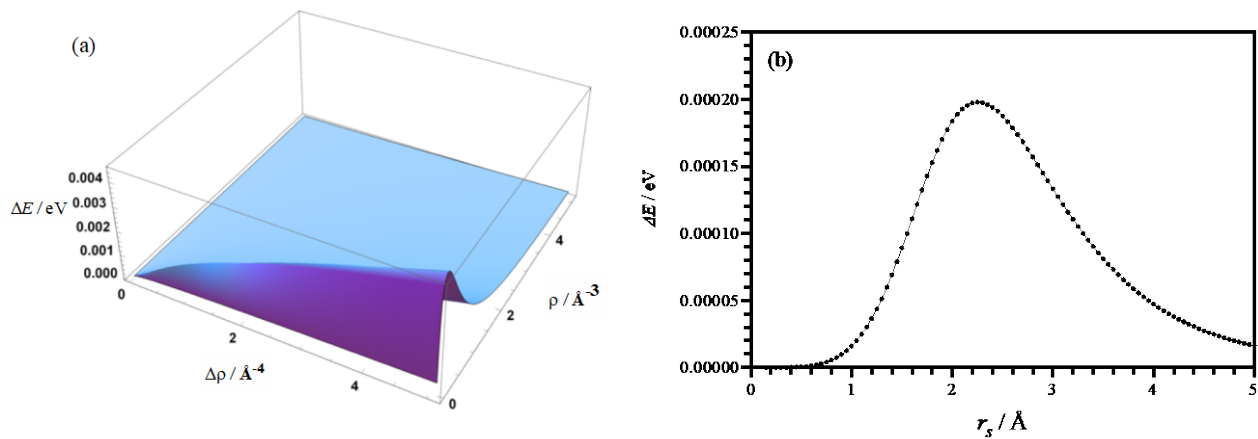

**Fig. 2.** The short-range exchange-energy difference $\Delta E = E_{xB88}^{SR\prime} - E_{xB88}^{SR}$ shown (a) as a function of the electron density $\rho$ and its gradient $\Delta\rho$, and (b) at $\Delta\rho = 0.1$ Å$^{-4}$ in terms of the Wigner–Seitz radius $r_s = \left(\frac{3\pi}{4\rho}\right)^{1/3}$.

To ascertain the significance of this error for calculated bond lengths and reaction energies of the model compounds from the ACCDB dataset, Table 5 shows the difference between results obtained using both methods with the GW and PBE pseudopotentials compared to results obtained using aug-cc-pV5Z. For both pseudopotentials, the MAD bond length errors are slightly larger when using the approximate method, as are the MAD reaction-energy differences. For the GW pseudopotentials, these are 0.0016 Å and 0.05 kcal mol$^{-1}$, respectively, and could be considered as negligibly small.

**Table 5.** Relative performance of the exact (Eqn. (14)) and approximate (Eqn. (20)) computational schemes for the short-range exchange contribution, applied to the test molecular data set.

| Property | Approximate | | Exact | |
|---|---|---|---|---|
| | GW | PBE | GW | PBE |
| Bond length MAD[a] (Å) | 0.0028 | 0.0044 | 0.0036 | 0.0048 |
| Reaction Energy MAD[a] (kcal mol$^{-1}$) | 0.51 | 0.63 | 0.34 | 0.59 |

a: compared to aug-cc-pV5Z results.

*3.5 Applications to materials.*

In Table 6, calculated lattice parameters are presented for the 3D materials: aluminum, graphene, diamond, silicon, NaCl, MgO, hexagonal boron nitride (h-BN), as well as results for a 2D h-BN monolayer. Both the GW and PBE pseudopotentials were used for CAM-B3LYP, and results are compared to both experiment and those from HSE06 calculations. The HSE06 results are found to be of useful accuracy, with a mean-absolute deviation (MAD) from experiment of only 0.014 Å. Nevertheless, CAM-B3LYP shows substantial improvement, reducing the MAD error to 0.0023 Å using the GW pseudopotentials (0.005 Å using PBE pseudopotentials).

**Table 6.** Comparison of predicted lattice parameters for CAM-B3LYP calculations (either GW or PBE pseudopotentials) for some iconic materials, compared to experimental and HSE06 results.

| Lattice | lattice parameter (Å) | | | |
|---|---|---|---|---|
| | Exp. | CAM-B3LYP/GW | CAM-B3LYP/PBE | HSE06 |
| aluminium | 4.0495[a] | 4.0503 | 4.050 | 4.022[h] |
| diamond | 3.5671[b] | 3.5671 | 3.566 | 3.548[h] |
| graphene | 2.464[c] | 2.4620 | 2.462 | 2.468[i] |
| silicon | 5.4310[a] | 5.4357 | 5.435 | 5.442[h] |
| NaCl | 5.6402[d] | 5.6472 | 5.654 | 5.662[h] |
| MgO | 4.217[e] | 4.2155 | 4.199 | 4.210[h] |
| 2D h-BN | 2.494[l] | 2.4948 | 2.492 | 2.488[j] |
| 3D[g] h-BN | 2.5044[f] | 2.5056 | 2.504 | 2.491[k] |
| MAD | | 0.0023 | 0.005 | 0.014 |

a: from.[75]
b: from.[76]
c: from.[77]
d: from.[78,79]
e: from.[80]
f: from.[81]
g: D3(BJ) dispersion correction used for CAM-B3LYP and HSE06 calculations.[82]
h: previous results using $\mu = 0.207$ Å$^{-1}$: Si- 5.435 Å, NaCl- 5.659 Å, diamond- 3.549 Å from.[83]
i: from.[84]
j: from.[85]
k: from.[86]
l: on Si/SiO,[87] other values 2.49 Å on Ni(111),[88] 2.50 ± 0.1 Å on Cu(111),[89] 2.50 Å on a metal-insulator-metal combination.[90]

Table 7 compares CAM-B3LYP/GW quasiparticle band gaps, lowest excitation energies, and exciton binding energies to observed data, $G_0W_0$, and $G_0W_0$-BSE calculations, as well as to results from HSE06 calculations. HSE06 is widely recognized as being the standard DFT method for use in calculating bandgaps,[91] with the $G_0W_0$ approach accepted as the best method useful for generally obtaining more reliable results for quasiparticle bandgaps, and similarly $G_0W_0$-BSE recognized for the calculation of optical excitation energies. Both $G_0W_0$ and $G_0W_0$-BSE are computationally very expensive in comparison to CAM-B3LYP and HSE06, limiting their application potential.

**Table 7.** Comparison of predicted quasiparticle band gaps and lowest optical excitation energies[a] for CAM-B3LYP/GW calculations for metallic, Dirac-cone, semiconducting, and insulating materials, compared to experimental, $G_0W_0$, $G_0W_0$-BSE, and HSE06 results.

| Lattice | Quasiparticle (orbital) band gap (eV) | | | Lowest optical vertical excitation energy (eV) | | | Electron-hole interaction energy (eV) | | |
|---|---|---|---|---|---|---|---|---|---|
| | $G_0W_0$ | CAM-B3LYP | HSE06 | Obs. | CAM-B3LYP | HSE06 | BSE/Obs.[q] | CAM-B3LYP | HSE06 |
| Al | | 0 | 0 | metal | | | | | |
| graphene | | 0 | 0 | Dirac cone | <0.003[b] | <0.0001[b] | | <0.003[b] | <0.0001[b] |
| diamond | 5.60[p] | 5.66 | 5.30[c] | 5.50[d] | 5.60 | 5.30 | ~0.1 | 0.059 | 0.004 |
| silicon | 1.29[p] | 2.26 | 1.16[c] | 1.17[d] | 2.24 | 1.16 | 0.015 | 0.022 | 0.001 |
| NaCl | 8.64[l] | 8.83 | 6.43[c] | 8.5-8.7[e] | 8.73 | 6.43 | ~0.1 | 0.099 | 0.002 |
| MgO | 7.7-7.8[m] | 7.54 | 6.46[c] | 7.67[d] | 7.46 | 6.45 | ~0.1 | 0.077 | 0.006 |
| 2D h-BN | 7.32[f] | 7.18[g] | 5.54[o] | 5.9-6.2[n] | 5.78 | 5.44 | 1.4 | 1.40 | 0.09 |
| 3D[h] h-BN | 6.48[i] | 6.62 | 5.58[k] | 6.08[j] | 5.91 | 5.57 | ~0.4 | 0.71 | 0.015 |
| MAD all | | 0.28 | 1.09 | | 0.30 | 0.76 | | 0.06 | 0.33 |
| MAD w/o Si | | 0.14 | 1.29 | | 0.15 | 0.91 | | 0.08 | 0.40 |

a: observed data is for spin-allowed transitions within the singlet manifold, calculated data is for the associated spin-forbidden transitions of marginally lower energy.
b: Dirac cone, evaluated for a 15 × 15 2D unit cell.
c: reported also:[91] diamond 5.34 eV, silicon 1.17 eV, NaCl 6.46 eV, MgO 6.45 eV.
d: observed.[92]
e: from.[93-95]
f: $G_0W_0$,[96] discussing also four previous calculations with results in the range 6.00 – 7.40 eV.
g: 7.61 eV using Eqn. (20) as an approximation to Eqn. (14).
h: D3(BJ) dispersion correction[82] used for CAM-B3LYP and HSE06 calculations.
i: $G_0W_0$ result, but this method incorrectly predicts a direct band gap[97] that originally received experimental support.[98,99]
j: from ref.[100] with a reorganization energy of 0.128±15 eV, that is assumed to dominate the exciton binding energy, based on a zero-phonon line at 5.955 eV and an indirect band gap.
k: also reported at 5.62 eV.[91]
l: $G_0W_0$ from.[101]
m: $G_0W_0$ refs.[102,103]
n: suspended on solids, refs.;[104,105] $G_0W_0$-BSE predicts 5.58 eV.[96,106]
o: previously reported values 5.56-5.69 eV depending on lattice parameters used.[107-109]
p: from.[110]

q: from available observed, $G_0W_0$, and $G_0W_0$–BSE results; for Si is shown the observed value.[111]

Aluminum is a metal, and all computational approaches used reproduce this property. Similarly, the Dirac cone in graphene is well reproduced, with the lowest optical excitation energy predicted by CAM-B3LYP to be < 0.003 eV (obtained for a finite-sized 15 × 15 supercell). Quantitative comparisons of properties are shown in Table 7 for the other materials. For diamond, the CAM-B3LYP/GW orbital bandgap is 5.66 eV, close to the $G_0W_0$ quasiparticle bandgap of 5.60 eV, with also the CAM-B3LYP/GW lowest optical transition energy of 5.60 eV being close to the observed value of 5.50 eV. Similar good agreement is found for the other medium-bandgap materials considered: NaCl, MgO, h-BN, and 2D h-BN.

Of all the materials properties considered, poor performance is obtained only for the quasiparticle bandgap and the lowest optical excitation energy in silicon: these quantities are overestimated by 1.1 eV. To put this result into context, Table 7 lists the MAD deviations for all materials studied, and the CAM-B3LYP error for silicon is noted to equal the determined MAD error for HSE06, whereas, excluding silicon, the MAD error for CAM-B3LYP is 9-fold lower than that for HSE06. Silicon is an extremely unusual material, leading to its unique and extremely important applications, with its bandgap arising as a result of near-cancellation of two primary chemical effects.[112] It is therefore quite difficult for computational methods to reproduce this property, leading to a history of methods developments in computational materials science that have silicon as a key reference marker. In contrast, no consideration of the properties of silicon was used in the design of CAM-B3LYP. Note that the calculated small exciton binding energy from CAM-B3LYP of 0.022 eV is close to the observed value of 0.015 eV,[111] whereas HSE06 depicts order of magnitude errors in this property. The generic description of silicon afforded by CAM-B3LYP would thus appear to be sound.

Comparing CAM-B3LYP band gaps and lowest excitation energies to HSE06 ones, Table 7 shows improvements of fourfold and 2.5-fold, respectively, using all data, increasing to 9-fold and 6-fold if silicon is excluded. In either case, the CAM-B3LYP exciton binding energies show fivefold improvement compared to HSE06.

It is not straightforward to discriminate between CAM-B3LYP results and those from $G_0W_0$ and $G_0W_0$–BSE as numerical differences between different reported $G_0W_0$ and/or $G_0W_0$–BSE results are mostly larger than the differences to CAM-B3LYP. For 2D h-BN, the CAM-B3LYP optical excitation energy is much closer to that observed than is the available $G_0W_0$–BSE result. From the preliminary results presented, it appears that CAM-B3LYP is competitive compared to both $G_0W_0$ and $G_0W_0$–BSE, but requires very much less computational resource and appears to be widely applicable throughout materials science.

## 4. Conclusions

We present an implementation of the CAM-B3LYP density functional into our copy of the VASP package. The implementation is tested by comparison to results obtained using atomic basis sets for 480 unique bond lengths and 1738 reaction energies as specified in the ACCDB benchmark database. Excellent agreement is obtained between the VASP implementation using plane-wave basis sets and the GW PAW pseudopotentials for the inner-core electrons and the all-electron atomic-basis set approach using the aug-cc-pV5Z basis. Use of PBE pseudopotentials rather than GW is found to be less accurate, but nevertheless useful in most applications. Also, increased values of ENCUT above the usual value for high-precision VASP calculations is found to be needed to get accurate results for compounds containing fluorine atoms. A rough summary of the comparison between results obtained using Gaussian basis sets and plane-wave basis sets is: PBE pseudopotentials give an accuracy comparable to 6-31+G* and cc-pVDZ, GW gives results similar to cc-pVTZ, and GW with ENCUT = 850 eV gives results similar to cc-pV5Z.

An approximate implementation of CAM-B3LYP was also considered in which equations used for the short-range exchange energy as part of the HSE06 specification. This was found to only degrade performance very slightly for calculated bond lengths and reaction energies. Nevertheless, for calculated band gaps in materials, it was found to lead to significant errors. Use of this approximation may expedient implementation of CAM-B3LYP into other plane-wave DFT codes.

Test applications of CAM-B3LYP to materials revealed results mostly in excellent agreement with experiment and high-level *ab initio* computational approaches such as $G_0W_0$ and $G_0W_0$–BSE; these results embody reduction of MAD errors by factors of 5-9 compared with results from the currently most widely used DFT functional for such applications, HSE06. One exception to this was found, pertaining to the band gap in bulk silicon.

Nevertheless, even for silicon, CAM-B3LYP produced order-of-magnitude improvement of the calculated exciton binding energy compared to HSE06, with a fivefold improvement in this property found considering all of the materials studied. The effect of charge transfer in simple materials like graphene, diamond, and silicon is expected to be minimal, owing to the very long-range, highly distributed, nature of the electronic transitions supported. Indeed, these materials were chosen for study as they could represent some of the worst-case applications for CAM-B3LYP. Yet no issues were found for graphene, a very unusual material, and CAM-B3LYP predicted considerably improved lattice vectors and exciton binding energies for diamond and silicon, as well as an improved quasiparticle band gap and an improved lowest optical excitation energy for diamond.

Turning now to the structured materials NaCl, MgO, and h-BN for which optical transitions involve localized charge-transfer processes that become delocalized over the extent of the crystals.

For these, the improvement of calculated properties using CAM-B3LYP compared to HSE06 is dramatic. Such systems form a paradigm for most materials. Improved spectroscopic properties for materials using CAM-B3LYP have already been reported.[54] A detailed study of the application of CAM-B3LYP to charge-transfer bands in h-BN defects will be presented elsewhere,[53] considering a situation in which HSE06 is inadequate to provide even a qualitative spectroscopic description.

**Acknowledgments**

This work was supported by resources provided by the National Computational Infrastructure (NCI), as well as Chinese National Natural Science Foundation Grant #11674212. Computational facilities were also provided by the ICQMS Shanghai University High Performance Computer Facility. Funding is also acknowledged from Shanghai High-End Foreign Expert grants to R.K. and M.J.F.

**Data availability statement**

Critical data is provided in Supporting Material, including equation derivations, *k*-point dependences, calculated bond lengths and reaction energies, and optimised molecular and materials coordinates. Additional information is available from the authors.